\documentclass[Phys]{SciPost}

\def\be{\begin{equation}}
\def\ee{\end{equation}}
\def\bea{\begin{eqnarray}}
\def\eea{\end{eqnarray}}
\def\ba{\begin{array}}
\def\ea{\end{array}}
\newcommand{\bes}{\begin{subequations}}
\newcommand{\ees}{\end{subequations}}

\def\CD{{\cal D}}

\begin{document}

\begin{center}{\Large \textbf{
An Infalling Observer and Behind the Horizon Cutoff
}}\end{center}

\begin{center}
Amin Akhavan,\textsuperscript{1 *}
Mohsen Alishahiha\textsuperscript{2  $\dagger$}
\end{center}

\begin{center}
{\bf 1}Young Researchers Club, Central Tehran Branch, Islamic Azad University, Tehran, Iran\\
{\bf 2} School of physics, Institute for Research in Fundamental Sciences (IPM)
P.O. Box 19395-5531,   Tehran,   Iran
\\
${}^*$amin$_-$akhavan@ipm.ir, ${}^\dagger$ alishah@ipm.ir 
\end{center}

\begin{center}
{\it Dedicated to Farhad Ardalan on the occasion of his 80th birthday}
\end{center}


\section*{Abstract}
{\bf
Using Papadodimas and Raju construction of operators describing  the interior of a black hole, we 
present a general relation between partition functions of operators describing inside and outside
the black hole horizon. In particular for an eternal black hole the partition function 
of the interior modes may be given in terms  those partition functions  associated 
with the modes of  left and right  exteriors. By making use of this relation we observe 
that setting a finite UV cutoff 
will enforce us to 
have a cutoff behind the horizon whose value is fixed by the UV  cutoff. The resultant 
cutoff is in agreement with what obtained in the context of holographic complexity.
}

\vspace{10pt}
\noindent\rule{\textwidth}{1pt}
\tableofcontents\thispagestyle{fancy}
\noindent\rule{\textwidth}{1pt}
\vspace{10pt}

\section{Introduction}\label{intro}
It is of a great interest to understand the physics behind the horizon, though its understanding 
might require full knowledge  of quantum gravity. 
Nonetheless,  AdS/CFT correspondence \cite{Maldacena:1997re}  has  provided a practical 
tool to explore, at least, certain features of the physics  behind the horizon.
Working within  the framework of  AdS/CFT correspondence  the problem may be 
rephrased as how to construct 
space time from boundary field theory. In this context holographic entanglement 
entropy \cite{Ryu:2006bv} was found useful to explore this possibility, though it might
 not be enough \cite{Susskind:2014rva}. 

Holographic complexity, by definition, is a quantity that is sensitive to regions behind the horizon.
Indeed, using either proposals for holographic complexity (``complexity=volume'' (CV) 
\cite{{Susskind:2014rva},{Stanford:2014jda}} or 
``complexity=action'' (CA) \cite{{Brown:2015bva},{Brown:2015lvg}})
one will have to deal with  a portion of space time located behind the horizon.
It is either a part of the Einstein-Rosen bridge, or a part of the Wheeler-DeWitt patch (WDW). 
Therefore one would naturally expect that the holographic complexity may have  information 
of the physics behind the horizon.

In  the CA proposal the late time behavior of complexity for 
an eternal black hole  is entirely given by the on shell action evaluated on 
the intersection of WDW patch with the future interior of the 
 black hole \cite{Alishahiha:2018lfv}. On the other hand one would expect that the late 
 time behavior of complexity is determined by physical charges, such as energy, which are 
 computed on the boundary of the space time. This fact may 
 indicate  a possibility of having a  relation between  inside and outside the horizon.
 We note, however, 
 that  this relation would result in an apparent puzzle as well.  
 While the physical charges are sensitive to  a UV 
 cutoff, the late time behavior of holographic complexity, giving entirely by interior of 
 the black hole, is blind to the UV cutoff.
 
 A remedy to resolve this puzzle was proposed in \cite{Akhavan:2018wla} (see 
 also \cite{{Alishahiha:2018swh},{Alishahiha:2019cib},{Hashemi:2019xeq},
 {Alishahiha:2019lng}}) in which it was argued that a UV cutoff will induce a cutoff 
 behind the horizon whose value is fixed 
 by the UV cutoff. 
  
 It is interesting  to see if the cutoff behind the horizon could also be seen 
 by other physical quantities. This is, indeed, the aim of this article to explore this possibility.
 To address this question one needs to look for an object that has a potential to probe physics
 behind the horizon. 
 
 To proceed, we note that, in the context of AdS/CFT correspondence,  
 there are several attempts to construct operators in the dual conformal field  describing the 
 interior of a black hole \cite{{Papadodimas:2012aq},{Verlinde:2013qya}}. Although there are
 serious concerns on the state dependence of these constructions \cite{{Bousso:2013ifa},
 {Marolf:2013dba},{Harlow:2014yoa}}, it is found useful to examine the cutoff behind the 
 horizon in this context. Of course  in what follows we will consider an eternal 
 black hole in which the situation is better understood.  Indeed, in most of our computations 
 we do not really need the construction of \cite{Papadodimas:2012aq} and the bulk 
 description of the operators is enough. 
 
This article is organized as follows. In the next section we will briefly review the construction 
of ``interior operators'' proposed in  \cite{Papadodimas:2012aq}. In section three we
shall consider partition functions of the interior and exterior  operators where we will also 
 present a relation between them. Then, we will examine this relation when the theory is
put at a finite cutoff.  Interesting  enough, we observe 
that setting a UV cutoff will automatically induce a cutoff behind the horizon whose value 
is exactly the one given by the holographic complexity.  The last section is devoted to discussions.
 

 \section{Interior operators}  
 
In this section we will review  the proposal  of \cite{Papadodimas:2012aq} to construct
conformal field theory operators describing the interior of a black hole. To fix our notation,  
we will consider an eternal black brane solution of an  Einstein gravity
 with negative cosmological constant. The action and the corresponding solution may 
 be written as follows
 \be\label{EHA}
S_{\rm EH}=\frac{1}{16\pi G}\int d^{d+2}x\sqrt{g}\left(R+\frac{d(d+1)}{L^2}\right),
\ee
and 
\be\label{BBS}
ds^2=\frac{L^2}{r^2}\bigg(-f(r)dt^2+\frac{dr^2}{f(r)}+d{\vec x}^2\bigg),\,\,\,\,
f(r)=1-\frac{r^{d+1}}{r_{h}^{d+1}}.
\ee 
The Penrose digram of the solution is depicted in the Fig. 1. 
\begin{figure}
\begin{center}
\includegraphics[width=.4\linewidth ]{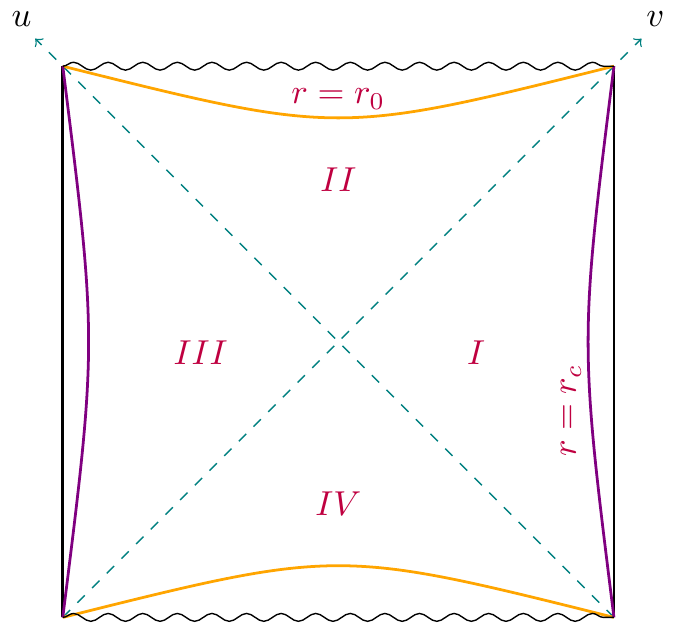}
\caption{Penrose diagram of an eternal black brane. The UV cutoff is denoted by $r_c$ and
the cutoff behind the horizon is denoted by $r_0$.}
\label{fig:gull}
\end{center}
\end{figure}
Here $\vec{x}$ are coordinates  parametrizing a $d$ dimensional flat space.
 Note that, there are also 
certain boundary terms which are needed in order to have a consistent variational principle.
 It is known that this model provides a gravitational description for a thermofield double state 
 \cite{Maldacena:2001kr}.
 
  Let us consider a generalized free field 
${\cal O}(t,\vec{x})$ in the field theory
side  whose  modes in the momentum space are denoted by ${\cal O}_{\omega,\vec{k}}$.\footnote{ Note that  by 
generalized free field one means that the correlators of the corresponding operators  factorize
at large $N$ limit \cite{ElShowk:2011ag}. } Then one may define a CFT (non-local) operator labeled by the AdS radial
coordinate $r$ as follows 
 \be\label{OI}
\phi^{I}_{\rm CFT}(t,\vec{x},r)=\int_{\omega>0}\frac{d\omega d^dk}{(2\pi)^{d+1}}\;
\bigg[{\cal O}_{\omega,\vec{k}}
f_{\omega,\vec{k}}(t,\vec{x},r)+ {\rm h.c.}\bigg].
\ee
When the function $f_{\omega,\vec{k}}(t,\vec{x},r)$ satisfies  the Klein-Gordon equation
for a free scalar in the back brane solution \eqref{BBS} with  normalizability condition  
near the boundary (and no condition at the horizon), 
this operator has the same correlators
 as that of a free-field propagating in the black brane background \cite{Papadodimas:2012aq}. 

This is, indeed,  a local bulk operator in region $I$ of the black barne  shown in the Fig. 1.
Similarly, one may define an operator associated with a free field associated with  the 
region $III$ as follows
 \be\label{OIII}
\phi^{III}_{\rm CFT}(t,\vec{x},r)=\int_{\omega>0}\frac{d\omega d^dk}{(2\pi)^{d+1}}\;
\bigg[\tilde{{\cal O}}_{\omega,\vec{k}}
{\tilde f}_{\omega,\vec{k}}(t,\vec{x},r)+ {\rm h.c.}\bigg].
\ee
Here we have used the ``tilde'' notation to make a distinction between operators describing 
 regions $I$ and $III$, though both of then are defined in the same dual conformal field theory.
 
To construct an operator describing the interior of the black brane (the region denoted by $II$ 
in the Fig. 1) one needs to consider both operators defined by ${\cal O}$ and 
$\tilde{{\cal O}}$. More precisely,  one has \cite{Papadodimas:2012aq} 
\bea\label{OBH}
\phi^{II}_{\rm CFT}(t,\vec{x},r)=\int_{\omega>0}\frac{d\omega d^dk}{(2\pi)^{d+1}}\;
\bigg[{\cal O}_{\omega,\vec{k}}\;
g^{(1)}_{\omega,\vec{k}}(t,\vec{x},r)
+{\tilde {\cal O}}_{\omega,\vec{k}}\; g^{(2)}_{\omega,\vec{k}}(t,\vec{x},r)+{\rm h.c.}\bigg],
\eea
where $g^{(1)}$ and $g^{(2)}$  still satisfy  the Klein-Gordon equation
for a free scalar in the back brane solution, of course with no boundary condition. 
One just needs to impose the continuity of the field at the horizon between regions $I$ and 
$II$ and also $III$ and $II$. 
For more  details  of how to construct  these operators 
the reader is refereed to the original paper \cite{Papadodimas:2012aq}.


\section{Partition function and cutoff}

In this section we would like to study possible  information one could get from 
the physics  behind horizon using  the operator given in the equation \eqref{OBH}
describing the interior of the black hole. Intuitively, from the construction of the 
operator \eqref{OBH}  one may extract following  information.

 First of all it seems that  in order to study  the region $II$ one needs twice the number of 
 modes as those in region $I$.  Secondly, since the operator \eqref{OBH} is a non-local 
 operator in the dual field theory whose non-locality parameter is given by the AdS 
 redial coordinate, imposing any restriction on the non-locality parameter
  (such as setting a UV cutoff) would restrict the range of the space time accessible 
  to those fields defined behind the horizon. 
 
 Clearly these facts  make a  connection between inside and outside the 
 horizon's physics, reminiscing  the phenomena observed in the complexity computations
 \cite{Akhavan:2018wla}.  Indeed, the aim of this section is to understand this connections better. 
 
 To explore a possible connection, let us study  the partition function of the  scalar field 
 we have considered in the previous section. It is worth noting that 
  since we are working within the context of AdS/CFT 
correspondence in which there is a relation between partition functions of gauge theory
and gravity, it what follows when we are computing the partition function we will not make 
an explicit distinction between these two pictures. Of course from the context, it should be
 clear whether we are using the partition function in the gauge theory or gravity  descriptions.

 To proceed, starting from the field theory description,  let us consider a {\it restricted}
  partition function in which  the integration is 
 taken over the fields  associated with 
 regions $I$ or $III$ of the corresponding eternal black hole\footnote{ A motivation of 
 considering this {\it restricted} partition function may be understood from the gravitional
 description of the model.  Indeed, starting from partition function in the gravity daul one may  
 approximate it by restricting our path integral over those field configurations satisfying 
the equations of motion (saddle point approximation) 
 \be
Z\propto \int  \CD \phi \, \,e^{-i S [\phi]}\,\delta({\rm e.\, o.\, m.})\,,
\ee 
where ${\rm e.\, o.\, m.}$ stands for ``equations of motion'' and therefore the path integral
should be performed for those fields given in the form of  \eqref{OI}, \eqref{OIII} or 
\eqref{OBH}.}
 \bea
Z^{(I)}\propto \int \CD\, \phi^{(I)} \, \,e^{-i S [\phi^{(I)}]},\;\;\;\;\;\;\;\;\;\;\;\;
Z^{(III)}\propto \int \CD\, \phi^{(III)} \, \,e^{-i \tilde{S} [\phi^{(III)}]}.
\eea
By making use of the explicit expressions of the corresponding fields in terms on 
${\cal O}$ and ${\tilde{\cal O}}$ given in the equations 
 \eqref{OI} and \eqref{OIII} one finds
  \bea
Z^{(I)}\propto \int_{\omega>0} \CD\,  {\cal O}_{\omega, \vec{k}}\;
 \CD\,  {\cal O}_{-\omega, -\vec{k}}\;\; e^{-iS[{\cal O}]},\;\;\;\;\;\;\;\;\;\;
Z^{(III)}\propto\int_{\omega>0}  \CD \,
{\tilde {\cal O}}_{\omega,\vec{k}}\; \CD \,
{\tilde {\cal O}}_{-\omega,-\vec{k}}\;\; e^{-i\tilde{S}[{\tilde {\cal O}}]}.
\eea
One may find  a similar expression for the field describing  the regions $II$ which, 
of course, should contain both sets of the  operators ${\cal O}$  and  ${\tilde O}$ 
 \be\label{PFII}
Z^{(II)}\propto\int_{\omega>0} \CD\,  {\cal O}_{\omega, \vec{k}}\; \CD \,
{\tilde {\cal O}}_{\omega,\vec{k}} \CD\,  {\cal O}_{-\omega, -\vec{k}}\; \CD \,
{\tilde {\cal O}}_{-\omega,-\vec{k}} 
\;\; e^{-i\bar{S}[{\cal O},{\tilde {\cal O}}]}.
\ee

 In general, the action appearing in the expression of \eqref{PFII} would be 
a complicated function of ${\cal O}$ and ${\tilde {\cal O}}$ and it cannot be decomposed in 
terms of actions for operators ${\cal O}$ and ${\tilde {\cal O}}$, separately;
$\bar{S}[{\cal O},{\tilde {\cal O}}]\neq S[{\cal O}]+\tilde{S}[{\tilde {\cal O}}]$. It is also important to 
note that a field in the region $II$,  is not a liner combination of those defined outside the 
horizon. Nonetheless 
for generalized free fields and taking  large $N$ limit in which  the 
corresponding correlation functions of the operators  factorize (see 
\cite{ElShowk:2011ag})
\be
\langle {\cal O}_1\cdots{\cal O}_ n {\tilde {\cal O}}_1\cdots
{\tilde {\cal O}}_m\rangle=\langle {\cal O}_1\cdots{\cal O}_ n \rangle\langle
{\tilde {\cal O}}_1\cdots
{\tilde {\cal O}}_m\rangle+{\cal O}\left(\frac{1}{N}\right)\,,
\ee
for any $n$ and $m$, one may write  
\bea
\frac{1}{Z^{(II)}}\frac{d^{n+m}Z^{(II)}}{dJ^nd\tilde{J}^m}\bigg|_{J=\tilde{J}=0}=
\frac{1}{Z^{(I)}}\frac{d^{n}Z^{(I)}}{dJ^n}\bigg|_{J=0}
\times\;\; \frac{1}{Z^{(III)}}\frac{d^{m}Z^{(III)}}
{d\tilde{J}^m}\bigg|_{\tilde{J}=0}+{\cal O}\left(\frac{1}{N}\right).\nonumber
\eea
Therefore for generalized free field and for large $N$, at leading order, one finds 
the following relation between partition functions associated with different regions
\be\label{ZZ}
Z^{(II)}\propto Z^{(I)} Z^{(III)}\,.
\ee
Actually this equation is a direct consequence of the equation \eqref{OBH} indicating that the
physical degrees of freedom in the region $II$ are constructed from those in two other regions.  
Note that for the case we are considering (eternal black hole) in which the operators 
${\cal O}$ and ${\tilde O}$ are identical and commuting one finds
\be
Z^{(II)}\propto (Z^{(I)})^2.
\ee

On the other hand using the fact that the partition function may be thought of as the effective 
action of the vacuum expectation value of the corresponding field when the  source
is set to zero \cite{Peskin:1995ev}, the above relation may be recast into the following form   
\be\label{G}
e^{i\Gamma^{(II)}[\varphi^{(II)}]}\propto e^{2i\Gamma^{(I)}[\varphi^{(I)}]},
\ee 
where $\varphi=\langle \phi \rangle$ and $\Gamma$ is the effective action. 
This is, indeed, an equation making  a connection between physics describing inside and out 
side the horizon.

Although we have found the equation \eqref{G} for a particular operator, in what follows 
we would like to consider the  effective action for the graviton in which the classical action 
is given by Einstein-Hilbert action \eqref{EHA}. In this case we will linearize the action around the 
classical solution \eqref{BBS} that  may be thought of as the expectation value of the graviton field.
Moreover,  since at the leading order the effective action is given by the classical action, 
we should  essentially compute on shell action for the solution \eqref{BBS} in different regions.

It is, however, worth noting that the equation \eqref{G} contains a  proportionality constant which 
should be fixed if one wants to extract concrete information. Actually,  since our ultimate aim is 
to compute the corresponding effective action (partition function)  for the case in which the 
theory is put at a finite cutoff \footnote{ Our motivation to consider the theory at finite a cutoff comes 
from ${\rm T{\bar T}}$ deformation of 
 conformal field theories \cite{{Zamolodchikov:2004ce},{Smirnov:2016lqw},{Cavaglia:2016oda}} 
   in which it was proposed that the corresponding holographic dual 
 may be provided by gravitational theories with a finite bulk radial cutoff
 \cite{{McGough:2016lol},{Taylor:2018xcy},{Hartman:2018tkw}}.
We note, however,  that there is a subtlety with Dirichlet boundary 
condition when we are dealing with the  gravity with a finite radial cutoff \cite{Witten:2018lgb}.
Therefore one should be careful when one wants to consider gravity with finite cutoff as a 
gravitational description for a ${\rm T{\bar T}}$ deformation. It is worth mentioning that 
there is another holographic description for the ${\rm T{\bar T}}$ deformation
 in terms of the mixed boundary condition\cite{Guica:2019nzm}. It seems
 that the ${\rm T{\bar T}}$ deformation may be holographically better understood in this 
approach.

 Although our main motivation comes from ${\rm T{\bar T}}$ deformation,
our aim was to study the system at finite cutoff which could set by hand. This might happen 
for example when we have an end of the world brane. 
 } 
one may compare two different cases in which the cutoff is or is not set to zero.\footnote{In the context of AdS/CFT correspondence the  partition 
function at a finite cutoff has been also studied in
 \cite{{Caputa:2019pam},{Ardalan:2019jnj}}.} Therefore, we are 
led to the following equation\be\label{G1}
e^{i(S^{(II)}_{\rm cutoff}-S^{(II)}_0)}=e^{2i(S^{(I)}_{\rm cutoff}-S^{(I)}_0)},
\ee
where the proportionality constant is dropped.\footnote{Note that changing the parameters (cutoff)
keeps the partition function intact and in the saddle point approximation the overall
proportionately constant is cutoff independent.}  Here $S^{(II)}$ and $S^{(I)}$ are on shell 
actions evaluated on the regions $II$ and $I$, respectively.

 
To proceed, let us  compute on shell action for the regions $II$ and $I$ when
the cutoff is set to zero.  To do so,  we note that the action of interest consists of several 
parts given by 
\be
S=S_{\rm EH}+S_{\rm GH}+S_{\rm CT},
\ee
where  $S_{\rm EH}$ is the Hilbert-Einstein term  given by the equation \eqref{EHA} and 
\bea
S_{\rm GH}=\frac{1}{8\pi G}\int d^{d+1}x\sqrt{-h}K,\;\;\;\;\;\;\;\;\;\;
S_{\rm CT}=\frac{-1}{8\pi G}\int d^{d+1}x\sqrt{-h}\frac{d}{L},
\eea
are Gibbons-Hawking and counter terms required to have a consistent action with 
a well defined variation principle that results in a finite free energy. It is, then, straightforward 
to compute on shell action for the solution \eqref{BBS} in the different regions. In particular for 
the  region $I$ one gets
\bea
S^{(I)}_{\rm EH}&=&\frac{\tau V_{d}L^d}{8\pi G}\left(\frac{1}{r_{h}^{d+1}}-\frac{1}{\epsilon^{d+1}}\right)\cr &&\cr
S^{(I)}_{\rm GH}&=&\frac{\tau V_{d}L^d(d+1)}{8\pi G}\left(\frac{1}{\epsilon^{d+1}}
-\frac{1}{2r_{h}^{d+1}}\right),\cr &&\cr
S^{(I)}_{\rm CT}&=&-\frac{\tau V_{d}L^d d}{8\pi G}\left(\frac{1}{\epsilon^{d+1}}-
\frac{1}{2r_{h}^{d+1}}\right),
\eea
so that
\be
S^{(I)}_0=\frac{\tau V_{d}L^d}{16\pi G} \frac{1}{r_{h}^{d+1}}\,.
\ee
Here $\tau$ is a cutoff in the time direction. On the other hand for the region $II$ one finds
\be
S^{(II)}_{\rm EH}=\frac{\tau V_{d}L^d}{8\pi G}\frac{-1}{r_{h}^{d+1}},\;\;
S^{(II)}_{\rm GH}=\frac{\tau V_{d}L^d(d+1)}{16\pi G}\frac{1}{r_{h}^{d+1}},
\ee
and the corresponding counter term vanishes. Therefore, one arrives at 
\bea
S^{(II)}_0=\frac{\tau V_{d}L^d(d-1)}{16\pi G} \frac{1}{r_{h}^{d+1}}\,.
\eea

Now, let us consider the case  in which we have a finite radial cutoff at $r=r_c$ that is 
associated with a UV cutoff in the dual field theory. It is then straightforward to compute  
the on shell action for this case.  Actually as far as the on shell action for  the region $I$ 
is concerned, the bulk part and  the  Gibbons-Hawking term of the action  have the same 
expressions except that one needs to replace $\epsilon\rightarrow r_c$. On the other hand 
from the explicit expression of the counter term one gets a non-trivial contribution. 
Indeed,  putting all terms together one arrives at 
\be
S_{\rm cutoff}^{(I)}=\frac{\tau V_{d}L^d}{8\pi G}\left(\!\frac{1-d}{2r_{h}^{d+1}}
+\frac{d}{r_{c}^{d+1}}\left(\!1-
\sqrt{1-\frac{r_{c}^{d+1}}{r_{h}^{d+1}}}\right)\!\!\right),
\ee
which  may be react into the following form
\be
S_{\rm cutoff}^{(I)}=\frac{\tau V_{d}L^d}{16\pi G}\left(\!\frac{1}{r_{h}^{d+1}}
+\frac{d}{r_{c}^{d+1}}\left(\!1-
\sqrt{1-\frac{r_{c}^{d+1}}{r_{h}^{d+1}}}\right)^2\right).
\ee

Form this expression it should be evident  that the on shell evaluated in the region $II$ 
cannot satisfy the equation \eqref{G1} unless we make a modification for the on shell action of
the inside  the horizon too. To proceed,  we will  assume that there is also a finite radial cutoff 
behind  the horizon located at $r_0$ preventing us to approach the singularity. With 
this assumption and taking into account all terms contributing to the on shell action
 one finds
\be
S_{\rm cutoff}^{(II)}=\frac{\tau V_{d}L^d}{8\pi G} \left(\frac{d-1}{2r_{h}^{d+1}}-
\frac{d}{r_{0}^{d+1}}\left(1-\sqrt{\frac{r_{0}^{d+1}}{r_{h}^{d+1}}-1}\right)\right).
\ee
Now using the on shell actions with and without cutoff and plugging them into the equation 
\eqref{G1},  one arrives at
\be
\frac{1}{r_{0}^{d+1}}\left(\sqrt{\frac{r_{0}^{d+1}}{r_{h}^{d+1}}-1}-1\right)
=\frac{1}{r_{c}^{d+1}}\left(\!1-
\sqrt{1-\frac{r_{c}^{d+1}}{r_{h}^{d+1}}}\right)^2,
\ee
that is exactly the same expression obtained in \cite{Akhavan:2018wla} relating the cutoff 
behind the horizon to the UV cutoff. In particular for a small radial cutoff ($r_c\ll r_h$) and  
at leading order the above equation reduces to
\be
r_0\,r_c^2\approx 2^{\frac{4}{d+1}}r_h^3.
\ee
Although it was not clear from the complexity computations  (see \cite{Akhavan:2018wla}) 
that why the cutoffs  $r_0$ and $r_c$ should come with different powers, it should  now be 
evident from the present consideration that it has to do with the fact that operators 
describing  behind  the horizon are constructed out of two copies of those describing 
outside the horizon.

 
\section{Discussions}

In this paper we have argued  that the partition function of the operators describing  
the interior of an eternal black hole is proportional to the product 
of partition functions of operators describing left and right exteriors of the black hole. 
Indeed, this is a direct consequence of the 
construction of interior operators in terms of those describing  the exterior regions  
proposed in \cite{Papadodimas:2012aq}. At leading order this connection  may be reduced
to a relation between on shell actions evaluated on the inside and outside the black hole. 

Using this relation we have computed on shell action for the solution \eqref{BBS} 
in different regions with the assumption that there is a  finite UV cutoff. 
We have observed  that setting  a finite UV cutoff enforces 
us to have a cutoff behind the horizon whose value is fixed by the UV cutoff. Interestingly 
enough, the expression we have found for  the cutoff behind the horizon  is the same as 
that obtained in the  context of holographic complexity.

In our computations we have assumed that the UV cutoff is the same for both left and right 
exterior regions of the corresponding eternal black brane. Nonetheless
one could also consider the case in which  the UV cutoffs for left and right regions are 
different. In the context of holographic complexity is was not clear how to proceed in such a
situation, though in the present approach it is straightforward  to deal with this case. 
Indeed, the only modification one needs to make  is to compute the 
on shell action of the two sides with different cutoffs. Doing so, one arrives at
\bea
\frac{1}{r_{0}^{d+1}}\left(\sqrt{\frac{r_{0}^{d+1}}{r_{h}^{d+1}}-1}-1\right)
=\frac{1}{2r_{c_L}^{d+1}}\left(\!1-
\sqrt{1-\frac{r_{c_L}^{d+1}}{r_{h}^{d+1}}}\right)^2
+\frac{1}{2r_{c_R}^{d+1}}\left(\!1-
\sqrt{1-\frac{r_{c_R}^{d+1}}{r_{h}^{d+1}}}\right)^2,
\eea 
where $r_{c_L}$ and $r_{c_R}$ are radial finite cutoffs associated with the  left and the right 
asymptotic regions. Note that in the limit of small  cutoffs, $r_{c_L}, r_{c_R}\ll r_h$,  one finds 
\be
r_0\, (r_{c_L}^{d+1}+r_{c_R}^{d+1})^{\frac{2}{d+1}}\approx 2^{\frac{6}{d+1}} r_h^3.
\ee
Of course, one may consider the case in which  one of the cutoff is approaching zero while the 
other one is kept finite. In this case one could still get a cutoff behind the horizon though its 
value at leading order  is grater  than the previous one by a factor of $2^{\frac{2}{d+1}}$.  

This situation  might be thought of as the case in which the corresponding solution 
represents a  typical black hole microstates of which the gravitational dual is provided by 
an eternal black hole when only a portion
of the left asymptotic region is taken into account \cite{deBoer:2018ibj}. Therefore we are 
left with a CFT at the boundary of the right exterior side and the left side is capped by a cutoff. 

In this case the operators describing the interior of the black hole are still constructed 
out of two copies of the exterior modes.  The second copy associated with the 
left side is given by the mirror operators \cite{Papadodimas:2012aq}. In this case 
one gets
\be
Z^{\rm interior}\propto Z^{\rm exterior} Z^{\rm mirror},
\ee 
leading to the following expression for the behind the horizon cutoff
\be
\frac{1}{r_{0}^{d+1}}\left(\sqrt{\frac{r_{0}^{d+1}}{r_{h}^{d+1}}-1}-1\right)
=\frac{1}{2r_{c_L}^{d+1}}\left(\!1-
\sqrt{1-\frac{r_{c_L}^{d+1}}{r_{h}^{d+1}}}\right)^2\,,
\ee
which for small cutoff and at leading order one gets $r_0\, r_{c_L}^{2}
\approx 2^{\frac{6}{d+1}} r_h^3$. 

It would be interesting to explore the role of the behind 
the horizon cutoff and its possible effects in other physical quantities.

\section*{Acknowledgements}
We would like to thank A. Mollabashi and K. Papadodimas for useful discussions. 
M. A. would also like to thank ICTP for very warm hospitality.   

\bibliography{SciPost_Example_BiBTeX_File.bib}


\end{document}